# Secure Data Transmission over Insecure Radio Channel in Wireless of Things (WoT) Network

Prokash Barman and Banani Saha

*Abstract*—Potential capacity of processors is enhancing rapidly which leads to the increase of computational ability of the adversary. As a result, the required key size for conventional encryption techniques is growing everyday for complex unbreakable security communication systems. The Public Key Cryptography (PKC) techniques which use larger keys cannot be fitted in tiny resource constrained Wireless of Things (WoT) devices. Some Symmetric Key Cryptosystems (SKC) use smaller keys, which can be fitted in the tiny devices. But in large networks where the number of nodes is in the order of $10^3$, the memory constraint does not allow the system to do so. The existing secure data communication in insecure medium uses various conventional encryption methods like Public Key Cryptography (PKC) and Symmetric Key Cryptosystems (SKC). Generally, modern encryption methods need huge processing power, memory and time. Also in some cases, Key Pre-distribution System (KPS) is used among different communicating devices. With the growing need of larger key size in the conventional secure communication system, the existing resources in the communicating devices suffer from resource starvation. Hence, the need of a novel mechanism for secure communication is inevitable. But the existing secure communication mechanisms like PKC, SKC or KPS do not ensure elimination of resource starvation issue in tiny devices during communication. In these existing conventional mechanisms, the plain text is generally converted into cipher text with greater size than the plain text at the device level, which leads to resource starvation. At the time of transmission, the cipher text at the device end requires more bandwidth than the plain text which puts bandwidth overhead on the broadcast channel (BC). To obliterate the resource starvation at the device end and on BC, we have proposed a Multi Channel Secure Communication mechanism in this paper. In the proposed method, utilisation of channel bandwidth and memory in tiny devices has been optimised as the size of both the plain text and the cipher text are almost same at the device level. We have developed the prototype of the proposed secured communication mechanism with the WoT devices for wireless transmission system where the ISM radio frequency band has been used as Broadcast Channel(BC). It has been observed that the adversary needs more iterations to identify the communication channels than the existing limit of security of value $2^{160}$ iteration. [24] [25]. Our proposed and implemented system is thus a robust, secured and resilient WoT communication mechanism.

## I. INTRODUCTION

IN the existing data communication, security is provided mainly with encryption in software level and sometimes also in hardware level which require huge computational power and time latency. In current scenario, the Wireless of Things (WoT) devices are deployed usually in remote environments with very little or no maintenance budgets for prolonged periods. Deployment such WoT devices with costly processors having very high power consumption, increases overall cost of the WoT networks and at the same time causes a maintenance requirement of changing batteries regularly. On the other hand, the exiting conventional secure data transmission systems have the drawback of huge computation complexity.

*a)* : To cope up with the growing need of secure data transmission in WoT environment, more advanced architecture is being adopted to increase the speed of data transmission or to reduce the overhead of securing data. However, with new developments in the hardware field, we face an adversary with equally powerful tools for breaking the security. In this paper we have proposed a novel multi channel secure communication mechanism of secure data transmission which imbibe into the process of data transmission channel. This ensures independence from the software and hardware encryption. The security of proposed WoT networks uses location based triangulation schemes and MAC authentication for verifying the authenticity of a node before sending data.

*b)* : The proposed method will have no complex key predistribution scheme(KPS) for establishment of links for nodes in a network. It uses the multi channel transmission mechanism, without any encryption or key establishment protocol. The implemented mechanism can also be used with any existing communication algorithm as our system only carries the data from the sender to receiver end as plain text with channel security mechanism. This leads to a robust heterogeneous secure communication. Various sections in this paper have been arranged as follows. Section-II deals with Existing Mode of Secure Communication. Section-III describes the Proposed Method of Secure Communication. The implemented prototype has been explained in section-IV. The Security of Proposed Method has been depicted in section-V. The Comparison between Existing and Proposed has been described in section-VI and section-VII concludes the paper.

## II. EXISTING MODE OF SECURE COMMUNICATION

The attacks of wireless communications can be classified as active and passive attacks. In case of active attacks, the attacker transmit jamming signals to disturb wireless communication. In passive attack the attacker eavsdrops the user message without disturbance. To prevent the eavesdropping, the transmitter is required to take account of leakage to the eavesdroppers in message transmissions [1]. Extensive work on secure wireless communication has been conducted in

Prokash Barman is working with the Department of Computer Science & Engineering, University of Calcutta, West Bengal,India, e-mail: pbcse_rs@caluniv.ac.in, Tel.: +91-7980208030

Banani Saha is working with the Department of Computer Science & Engineering, University of Calcutta, West Bengal,India, e-mail: bsaha_29@yahoo.com, Tel.: +91-9433130715 point-to-point communications. The work of Shannon has introduced the concept of information theoretic secrecy [2]. The Shannon's concept has been generalised in [3] to a fundamental network namely wiretap channel. (see figure 1).

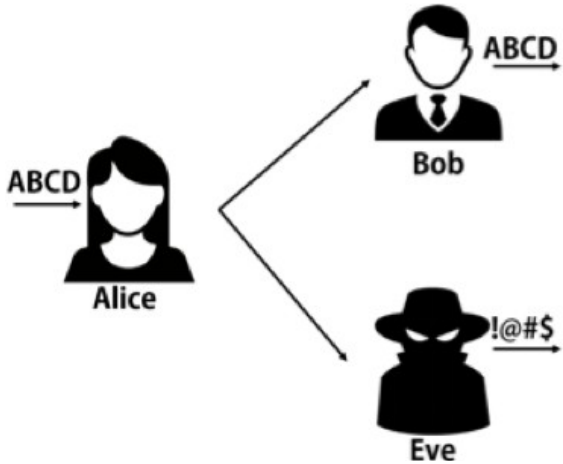

Fig. 1. Structure of basic wiretap channel

The wiretap channel has transmitter (Alice), Receiver (Bob) and Eavesdropper (Eve). The link between Alice and Bob is the **main link**. Where as link between Alice and Eve is called **wiretapping link**. The **Signal to noise ratio (SNR)** of Bob must greater than that of Eve for Secure communication in wiretap channel. To overcome the problem different techniques such as multi-antenna transmission [4]-[6] and artificial noise transmission [7] have been proposed.

In case of multi user scenario, secure communications have been considered in broadcast channels (BC) with eavesdroppers. In this BC a transmitter sends messages to multiple legitimate users without any information leakage to eavesdropper [8]. The BC with eavesdroppers are classified into two types- the **wiretap BC** [8] and the **BC with confidential messages** [9] [10]. The wiretap BC deals with the communication security problems in the network with external eavesdroppers. Since the external eavesdroppers have no interaction with any terminal in the network, the **channel state information (CSI)** about the eavesdroppers is generally assumed to be unknown at the transmitter. For this reason, it is impossible to always guarantee the communication security in wiretap BC. Most work on wiretap BC has considered transmission strategies for guaranteeing security with a high probability instead of guaranteeing the security all the time. [11]- [13]. With the imperfect or no CSI of the eavesdroppers, transmitting artificial noise onto the null space of the legitimate receiver's channel is considered as an effective way to improve secrecy outage capacity or secrecy outage probability by exploiting multiple transmit antennas. Since all the receivers except the legitimate receiver are affected by the artificial noise, the artificial noise transmission can degrade the link condition of eavesdroppers even though the global CSI is unavailable. As the second type of channel, the BC with confidential message deals with the communication security problems in the network with internal eavesdroppers. In other words, the transmitter assumes all the internal users can potentially act as eavesdroppers in transmission of confidential messages. The global CSI is generally assumed in this channel model since the internal users have some interaction with the transmitter contrary to the passive external eavesdropper in wiretap channels. [14]

### A. Conventional Cryptography Methods

Encryption of information is alternate way out to protect the network and data from attack, which is widely used and also popular. The most common and popular algorithms used for encryption are RSA, ECC, AES, 3DES, MD5 and SHA authentication. These encryption system are heavily computational. [15]. For each potential message, a specific code is used to check the validity of the message. By using protocols such as IPSec, the accessibility and authenticity will be provided for the data flow. For implementing these algorithms, there should be specific and dedicated processors such as Digital Signal Processors (DSP) to provide the required highly computational process. In most cases, these processors have only one layer of encryption algorithm. Security should be provided in multi layer for robust security. Conventional security in IoT/ M2M is viewed in forthcoming sub section.

### B. Security in IoT/M2M

As the applications of the IoT/M2M affect our daily lives, whether it is in the industrial control, transportation, Smart Grid or healthcare verticals, it becomes imperative to ensure a secure IoT/M2M system. With continuous change of IP networks, IoT/M2M applications have already become a target for attacks that will continue to grow in both magnitude and complexity. The scale and framework of the IoT/M2M make it a persuasive target for companies, organizations, nations, and more importantly people hackers. The targets of attack are abundant and cover many different industries. The possible impact could be reflected in minor irritant to significant damage to the infrastructure and loss of human life.

The device identity and mechanisms to authenticate the IoT is one of the fundamental elements in securing an IoT infrastructure. Many IOT devices may not have the required computation power, memory or storage to sustain the current authentication protocols. There are various strong encryption and authentication schemes are used today, which are based on cryptographic suites such as Advanced Encryption Suite (AES) for confidential data transport, Rivest-Shamir-Adleman (RSA) which is used in digital signatures and key transport and Diffie-Hellman (DH) for key negotiations and management. As the protocols are robust, they require high compute platform, a resource that may not exist in all IoT-attached devices. Therefore, authentication and authorization will require appropriate re-engineering to accommodate our new IoT connected world.

TABLE I
KEY SIZE AND KEY SIZE RATIO BETWEEN ECC AND RSA/DSA.

| ECC Key Size | RSA/DSA Key Size | Ratio (ECC/RSA) |
|---|---|---|
| 106 | 512 | 1:5 |
| 132 | 768 | 1:6 |
| 163 | 1024 | 1:7 |
| 192 | 1536 | 1:8 |
| 210 | 2043 | 1:10 |
| 256 | 3024 | 1:12 |
| 384 | 7680 | 1:20 |

TABLE II
DNA NUCLEOTIDE TO BINARY CONVERSION TABLE.

| DNA Nucleotide Base | Binary equivalent |
|---|---|
| Adenine (A) | 00 |
| Thymine(T) | 01 |
| Guanine(G) | 10 |
| Cytosine(C) | 11 |

These authentication and authorization protocols require user intervention in terms of provisioning and configuration. However, many IoT devices will have limited access, thus required initial configuration to be protected from theft, tampering and other forms of compromise throughout its usable life, which in many cases could be years.

These issues can be overcome with new authentication schemes which can be built using the experiences of today's strong encryption/authentication algorithms. New technologies and algorithms are being worked on. As an example, the National Institute of Standards and Technology (NIST) has chosen the compact SHA-3 as the new algorithm for the so-called "embedded" or smart devices that connect to electronic networks but are not themselves full-fledged computers. [17]

*C. DNA encoded ECC for IoT security*

To ensure multi layer security in IoT DNA encoded ECC method is proposed in [23]. Elliptic curve cryptography (ECC) was discovered by Neal Koblitz and Victor Miller in 1985 [18] [19]. ECC is the most efficient public key encryption method. To enhanced cryptographic complexity in ECC, the concept of elliptic curve is used. Generally, Elliptic Curve Cryptography is used to compare with the public key encryption systems like RSA and diffie-hellman key exchange problem. ECC helps to provide highest security in low power consumption IoT / WoT devices. Some public key encryption systems like RSA, D-H key exchange and Digital Signature Algorithm (DSA) are suitable for devices with high computation ability. But IoT /WoT or cloud computing low computational powered devices could only be fitted with minimal computational encryption methods. [16]

The ratio of ECC and RSA/DSA key size is described in Table-I by Bafandehkar, Mohsen & Yasin, Sharifah & Mahmod, Ramlan & Zurina, Mohd Hanapi. (2013) [20]. In the Table-I it is seen that ECC has lowest key size compared to RSA/DSA cryptography methods. This imply that ECC may also be suitable to fit into the tiny IoT/ WoT devices. Need of DNA encoding mechanism over ECC is depicted in next sub section.

*1) Need of DNA encoded ECC in IoT :* To protect private communication and keep it private from everyone except the intended recipients cryptography is the used. To securing IoT devices and many other electronic products proper use of cryptographic technique is essential. In the blog titled, "Robust IoT security costs less than you think," shown that the effectiveness of IoT security can be accomplished by a cryptographic IC that adds less than $1 to an IoT device bill of material (BOM). Implementing public-key cryptosystem for IoT devices requires significant expertise, but organizations may hire IoT security partners who are capable of providing complete, robust security solutions, typically in a matter of weeks, that is generally performed in parallel with different development efforts.

Obviously, any key will in theory be broken by a brute-force attack with sufficient computing power. The practical approach of contemporary cryptography is to use a key of adequate enough length that it can't be broken without an extraordinary amount of computing power that would be significantly more than the value of the contents that the cryptosystem protects. In the above mentioned sub-$1 cryptographic IC utilizes 256-bit Elliptical Curve Cryptography (ECC) keys, which is so secure that the computational power to break a single key may require computing resources equal in cost to 300 million times the entire world's annual GDP (78 trillion USD) working for an entire year.

IoT product maker has no valid excuse to continue to ignore IoT security risks because the modern cryptographic ICs make IoT security most affordable.

*2) DNA Encoded Elliptic Curve Cryptography Scheme:* The elliptic curve cryptography and DNA cryptography are the most modern cryptographic technique. Thus, for better security, characteristics of both the systems may be combined to construct highly secured cryptographic techniques(as proposed in [23]). In the new system for encoding of plain text we propose ( [23])to use the insertion method of DNA cryptography as used in [21].

The plain text is converted to its equivalent ASCII value. Then the ASCII values are converted into binary. From publicly available sequences of DNA, we choose known DNA nucleotide sequence. Both the sender and the receiver should synchronize with the chosen DNA sequence. The DNA nucleotide sequences are converted into binary using Table-II.In this stage we will obtain several pairs of binary numbers. All the binary number pairs are concatenated to make a long binary number. Then the obtained binary number is broken into several segments. Here an arbitrary number of bits, greater than 2 will be taken in each segment. Now each bit of converted binary plain text inserted into the beginning of the binary segments of nucleotide sequence. The segments are concatenated again and converted to Nucleotide letter. (A,T,G, C). Now the new sequences are converted into decimal following conversion rules in Table-II. [22]

*a) :* In the proposed system [22] the above steps are used for encoding. For encryption the decimal numbers are converted into elliptic curve point using Koblitz method [21]. This point is called plain text point. The points are encrypted into another elliptic curve point using ECC encryption expression (1). ECC encryption process is performed with the help of its generated keys. The encrypted points on the elliptic curve are called cipher text points. This point is sent to the receiver. In the reverse process, the receiver will derive the plain text message.

kG, Pm + kPB ——————————(Expression-1)

Pm + kPB - nB (kG) = Pm + k (nB) G - nB (kG) = Pm ——————————(Expression-2)

Where,
G - *Generated Points*
Pm - *Plaintext points*
k - *Random number chosen by user*
PB - *Public key of another user*

The cipher text points are deciphered using ECC decryption expression (2). Deciphered points are converted into numbers using Koblitz's method. These numbers are decoded to an unknown DNA nucleotide sequence. From the unknown DNA sequence, known DNA nucleotide sequence (S) is decoded to obtain plain text.

### III. PROPOSED METHOD OF SECURE COMMUNICATION

In the conventional secure communication, data encryption is performed within the communicating devices. The network graph for the devices has to be pre established for such communication. For pre establishment of network graph, secure link and secure connectivity are necessary [26]. Pinaki et al. depicted the procedure of offline generation of keys and different methods of pre-distribution keys among the devices to be connected in a network graph [26]. In the Key Pre-distribution System (KPS), encrypted keys are distributed among devices before establishment of a network graph. Hence, the communicating devices need to generate offline keys and encrypt them for transmission to other devices within network. This leads to computational overhead at the device level and bandwidth utilisation is maximised before establishment of network Signed Weighted Graphs [26]. In this paper a new Multi channel transmission mechanism has been proposed to secure the communication channel. In our method, the entire bandwidth of a communication channel is divided into several channels. Plain text is also sliced into multiple parts, each part of the plain text is then transmitted through randomly selected channel. The intended receiver reconstructs the sliced plaint text. The process has been depicted in Figure - 2. In the system pre establishment of network graph and encrypted key pre distribution are not necessary. So, processing overhead and bandwidth utilisation have been optimised in our proposal. The communication strategy of the system has been described in detail in section-III-A.

#### A. Communication of Data between devices and network

The Figure-2 depicts the communication strategy adopted in this paper. Each transceiver is constructed with one master tiny device and multiple tiny slave devices. The procedure for the basic operations on input data by the master device is as follows:

*Input: Plain text (P)*
*Output: Randomly allocation of converted parts of P (R') to slave tiny devices*

```
Procedure
  Begin
    Input plain text P
    Convert P into hexa decimal  P'
    Apply ROT13 on P', get R=ROT13(P')
    Apply secret sharing on R, get R'
    Break R' into parts, R'1, R'2 ...R'n
    Randomly allocate parts of R' to tiny slaves
  End
end Procedure
```

The slave devices take part in actual data transmission in wireless network. The I2C protocol is adopted for intra-device communication within transceiver. The SPI protocol is used to communication between transceivers. We have used the ISM (Industry, Scientific and Medical) band as BC for implementation of our proposed secure communication mechanism. The available transmission frequency range in ISM band is divided into multiple small channels and allocated randomly to each slave tiny device for transmission. The ISM Band has been narrated in sub section -III-A1. The I2C protocol is described in sub section-III-A2. The ROT13 encryption method is elaborated in sub section-III-A3. In sub section- III-B we have described the method of the proposed secure communication.

*1) ISM Band:* The ISM radio bands are portions of the radio spectrum reserved internationally for industrial, scientific and medical (ISM) purposes other than telecommunications [27]. Examples of applications for the use of radio frequency (RF) energy in these bands include radio-frequency process heating, microwave ovens, and medical diathermy machines. The powerful emissions of these devices can create electromagnetic interference and disrupt radio communication using the same frequency, so these devices are limited to certain bands of frequencies. In general, communications equipment operating in these bands must tolerate any interference generated by ISM applications, and users have no regulatory protection from ISM device operation. [28]

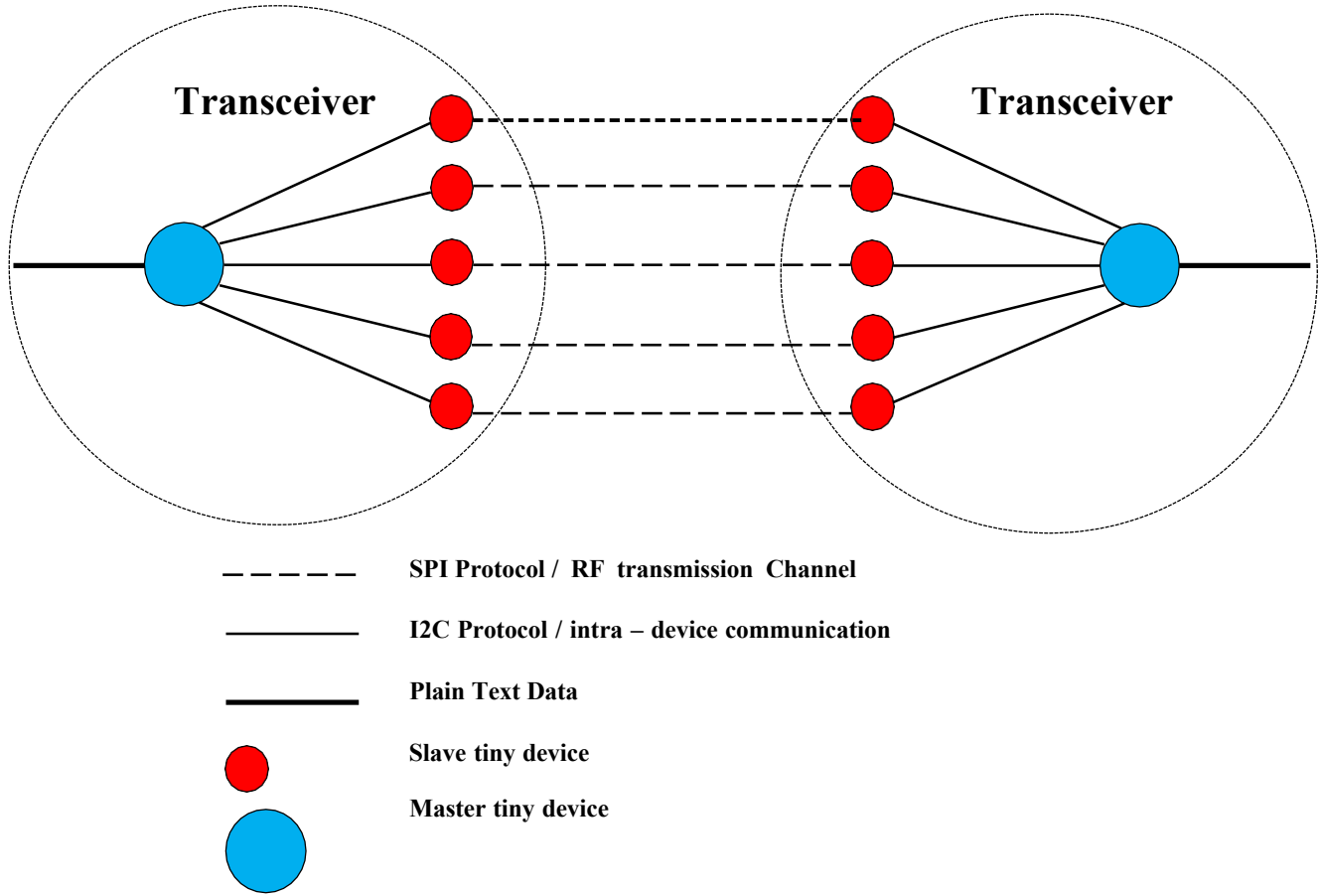

Fig. 2. Communication of Data within devices and in a network

*a) Definition:* The ISM bands are defined by the ITU Radio Regulations (article 5) in footnotes 5.138, 5.150, and 5.280 of the Radio Regulations. Individual countries' use of the bands designated in these sections may differ due to variations in national radio regulations. Because communication devices using the ISM bands must tolerate any interference from ISM equipment, unlicensed operations are typically permitted to use these bands, since unlicensed operation typically needs to be tolerant of interference from other devices anyway. The ISM bands share allocations with unlicensed and licensed operations; however, due to the high likelihood of harmful interference, licensed use of the bands is typically low. In the United States, uses of the ISM bands are governed by Part 18 of the Federal Communications Commission (FCC) rules, while Part 15 contains the rules for unlicensed communication devices, even those that share ISM frequencies. In Europe, the ETSI is responsible for regulating the use of Short Range Devices, some of which operate in ISM bands. [28]

*b) Selection of Frequency for BC:* The allocation of radio frequencies is provided according to the Article 5 of the ITU Radio Regulations (edition 2012) [29]. We have choose 2.4

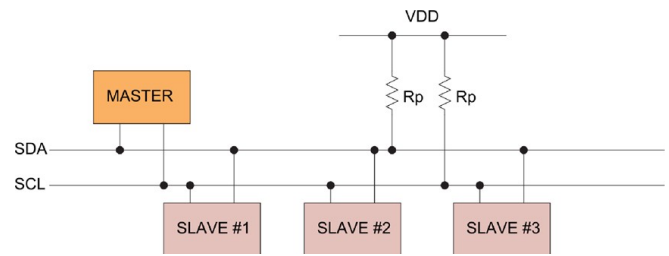

Fig. 3. I2C Multimaster Multislave Configuration with Pull up Resistors [30]

GHz-2.5 GHz ISM frequency range for the BC of our proposed system. The selected frequency range has a bandwidth of 100 MHz which may divided into $10^5$ distinct channels with 1 kHz frequency in each channel.

*2) I2C Protocol:* There are many protocols for transmission and reception of data from one device to another device, but I2C (Inter-integrated Circuit) is a simplest protocol because it has only two wire lines SDA (Serial Data Line) and SCL (Serial Clock Line) and it has multi-master capability unlike SPI (Serial Peripheral Interface) protocol and also addressing of I2C is simple because it does not require any CS lines used in SPI and it is easy to add extra devices on the bus [30] - [35]. I2C is a Synchronous Protocol unlike UART (Universal Asynchronous Receiver/Transmitter).There is so many application of I2C protocol in real time such as micro controller, wireless communication and many more [32] [34] [35].

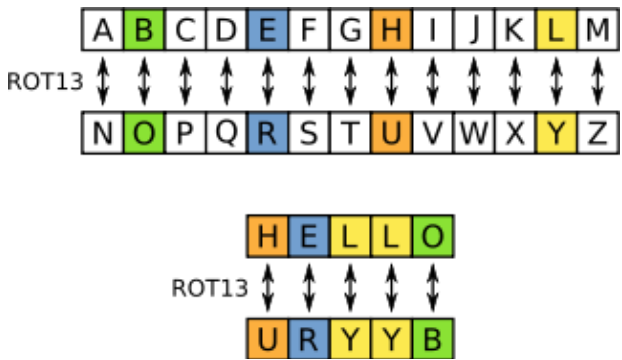

Fig. 4. ROT13 replaces each letter by its partner 13 characters further along the alphabet. For example, HELLO becomes URYYB (or, conversely, URYYB becomes HELLO again). [37]

*a) :* Both SDA (Serial Data Line) and SCL (Serial Clock Line) are bidirectional that provides a simple and efficient communication between devices. In I2C we can connect multiple masters and multiple slaves but single master and multiple slaves combination mostly used. Master is the device which generates clock signal, which initiates a transfer and which terminates the transfer. Slave is the device which addressed by a Master. [30] [31] [33].

*b) :* With the minimal circuit design I2C protocol efficiently interconnect multiple devices. Hence for the prototype of our proposed mechanism we have choose Inter integrated Circuit (I2C) protocol for intra-device communication.

*3) ROT13 Encryption:* A variety of Caesar's cipher called ROT13 is still used by some software and involves a shift by 13 characters: P = ROT13 (ROT13 (P)), so encrypting text with ROT13 twice gives you the original text. [36]

Applying ROT13 to a piece of text merely requires examining its alphabetic characters and replacing each one by the letter 13 places further along in the alphabet, wrapping back to the beginning if necessary . A becomes N, B becomes O, and so on up to M, which becomes Z, then the sequence continues at the beginning of the alphabet: N becomes A, O becomes B, and so on to Z, which becomes M. Only those letters which occur in the English alphabet are affected; numbers, symbols, white space, and all other characters are left unchanged. Because there are 26 letters in the English alphabet and 26 = 2 x 13, the ROT13 function is its own inverse [37].

### B. Proposed Secured Communication Method

*1) Sending Of Data:* The process of sending data has been depicted in Figure-5 and elaborated below :

1. The master device gets plain text from upper layer of the node.
2. The plain text in converted to hexadecimal format for ease of binary calculations. Let's call this **hex string digest**.
3. A simple **ROT13** cipher is applied to the digest and converted into cipher text.
4. The cipher text is then sent over to the secret sharing module which receives data from two registers N and K. [Figure-7] In this module the cipher text is broken into N pieces out of which at most K pieces are necessary to build the entire message. Let's call these pieces bits.
5. Simultaneously for each bit a random number is generated using a linear feedback shift register [Figure-7]. These numbers are be mapped to a frequency between 2.4 GHz- 2.5 GHZ [Sub section-III-A1]. This allots a unique frequency for a bit in one session.
6. After all bits are created and all bits have received their respective frequencies. All N bits are sent to slaves for transmission.

The first 5 steps of the mechanism are completed in the master tiny device of a transceiver. The slave tiny device of the transceiver only transmit the bit wise data over the assigned frequency.

*2) Receiving Of Data:* The process of receiving data has been depicted in Figure-6 and elaborated below :

1) The receiver uses a predefined value n(same for the entire network) to start the process.
2) The seed value from the seed register is sent to the linear feedback shift register (LFSR), which generates n random values. The important point to be noted is that the LFSR generates identical values at the sender and receiver side at a given time t = T if we ignore the clock skew.
3) For every generated value from the LFSR, we get the exact frequency value from the frequency lookup table.
4) These frequencies are in turn sent to the slaves, which open reading pipes for these given frequencies.
5) Once the data is received at all the reading pipes the data is sent to the master, where the processing of the data is taken place.
6) The first component of this processing is to order the data by their respective sequence numbers followed by reconstruct the message via secret sharing.
7) Then the content is decrypted using the ROT13 cipher used for encryption[see section-III-A3]. Lets call this message as plain-text.
8) The plain-text is then finally converted to the standard ASCII format and sent to the upper layers for further processing.

An important component in this process is the packet structure. The packet must contain certain identifiers, that can be used to pinpoint the location of origin. The packet consists of 4 fields.The first bit of the packet is 1. The next 8 bit field is reserved for putting a sequence number of the packet and for providing the count of the number of packets sent from a given sender to a given receiver. The next 32 bits provide the device id of the sender. The next 80 bits are reserved for data in hex format. So roughly each packet allows us to send 10 bytes of data. Since the communication is parallel, we get 10n bits of data where n is the number of channels used.

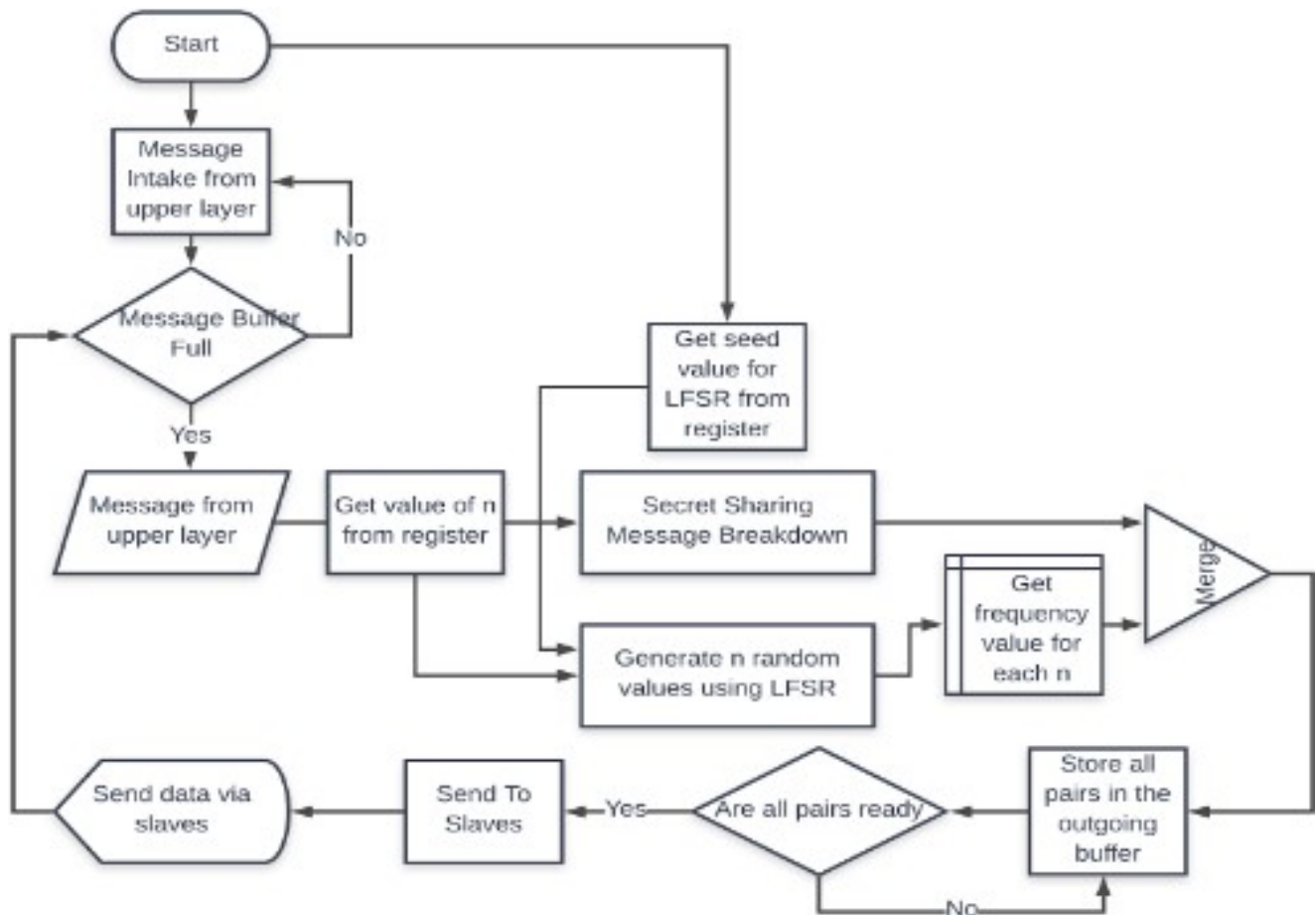

Fig. 5. Proposed Sending Method

Moreover since the system works on clock synchronization, we need some measure to take into account the clock skew among various nodes present in the network. To solve this problem we have implemented a synchronization scheme to be used after a fixed interval of time. There is a special packet called a SYNC packet, which is to be flooded by all nodes at t = T. The value of T is determined by the producer of the devices after considering the minimum admissible clock skew. The SYNC packet structure is stated below.

1) The first bit is 0.
2) Next 32 bits are reserved for the node ID.
3) Next 32 bits are reserved for the local clock count.

Before the explanation let us consider two concepts

1) Local Time : Local time is the count of clock pulses as seen by each device in the network. For each device the local time in the network maybe different.
2) Global Time : This is the count of clock pulses by a clock maintained outside the system. Global time remains same for all nodes in the network.

The sequence of events a $T = T_{max}$ where T is the local time of a node

1) At $T = T_{max}$ the nodes which have reached the given milestone, will flood out the SYNC packet over all channels.
2) It is assumed that in an n node network every node will receive all n-1 packets.
3) From this set of packets all nodes will pick the maximum of all the local clock counts and set it as their local time
4) It will send out a reply containing the device id of the packet that was used to set its time.

IV. IMPLEMENTED PROTOTYPE

The prototype of the transceiver is designed in figure 7. The transceiver has different modules as described below.

1) System Clock : A clock pulse generator provides a sense of time to the node and the sender.
2) Secret Sharing Module : This is a block that gets the message in hexadecimal format and converts this message into a series of N (provided by user) floating point numbers that can be used to reconstruct the message
3) Master Buffer : A storage space for all the small pieces of the message after it is broken down and before being sent to the slaves.

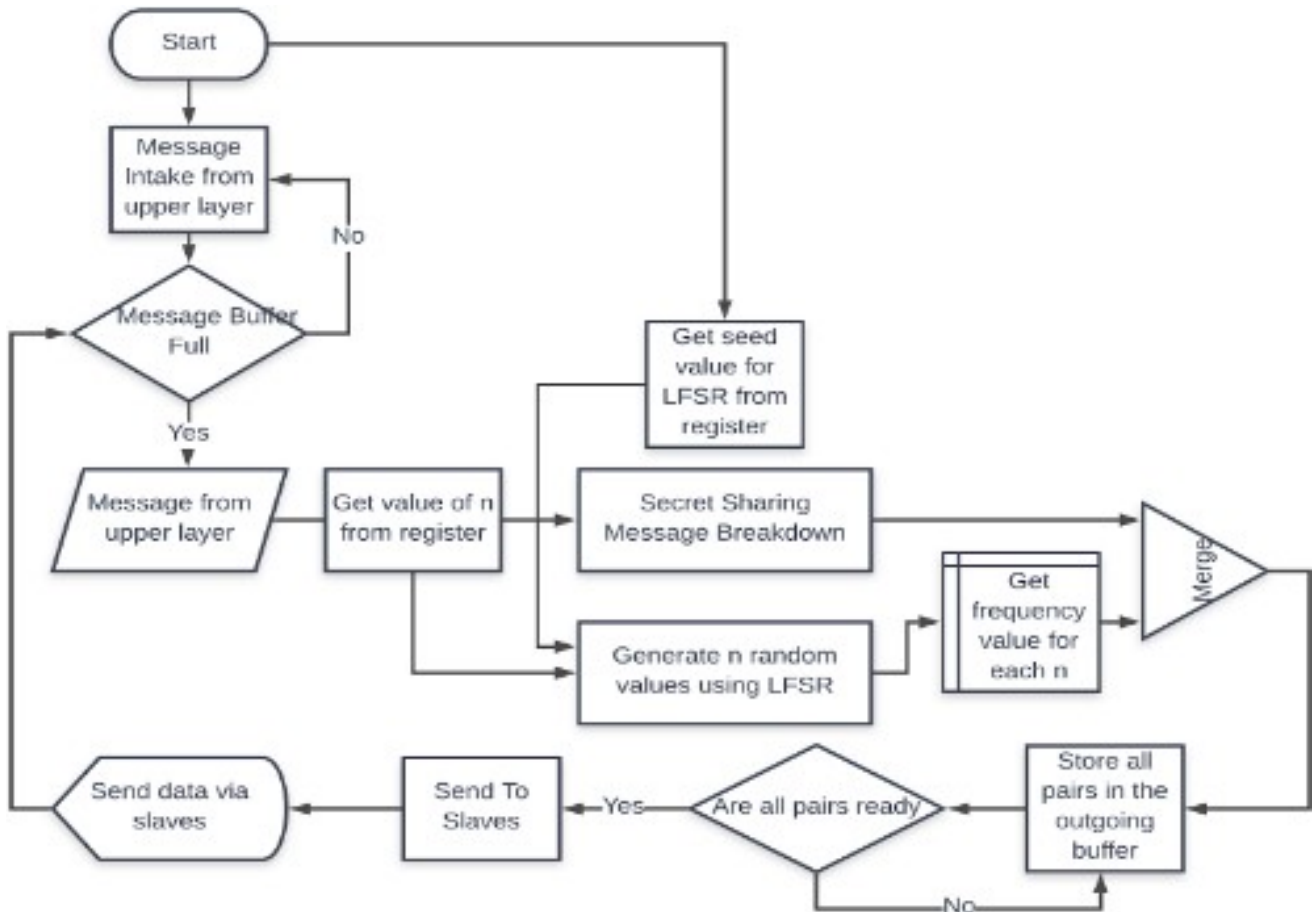

Fig. 6. Proposed Receiving Method

4) **Ready Flag** : A Boolean variable that represents if the master is ready to forward data to the slaves.
5) **Linear Feedback Shift Register** : A hardware system that generates a random integer for every clock pulse based on a seed value provided by the user.
6) **Frequency Lookup Table** : A hardware unit that holds a mapping from integers to frequency values. This system is basically a multiplexed memory unit.
7) **Master Outgoing Buffer** : A buffer that holds pairs of values, where one value is the message part and the other is the frequency at which it is to be transmitted.
8) **Data Receiver Logic** : This logic is present on the slave side and receives data from the data bus of the master. It is the responsibility of this module to separate data and frequency values and store them in their respective buffers.
9) **Transmission Hardware** : This part of the slave module deals with the specifics of transmitting the data on the provided frequency. On successful transmission this block switches on a success flag in itself.

## V. SECURITY OF PROPOSED METHOD

The principal claim of security comes from the point that the data is being in a parallel fashion over the channels. The channels are taken k at a time out of a huge number of channels, say n. This means the adversary has to choose $\binom{n}{k}$ where n is the total number of channels available and k is the number of channels being used. In our case the channel count is $10^5$ and the number of channels used per communication depends on the user but must be between 5 and 10 because at least 5 channels are needed to guarantee $2^{128}$ bits of security and less than ten to reduce processor load. Moreover it can be considered that if the adversary captures at least one packet among the sent packets, it can get some information about the total nature of data. To prevent this we are using a technique called finite field secret sharing which uses a modulo p polynomial where p is a prime number. This reduces the probability of getting complete information about the packets.

**Information Capture** : Information capture is the event when, at least one bit of the packet is correctly decrypted by the adversary.

Let e be the event of information capture, then, without secret sharing, let the probability of data being captured be $P_1$. Since each packet reveals some information about the entire data. Even when one packet is captured some information is revealed. Hence the probability $P_1$ accounts for capture of at least one or more , up to all k partitions of the data being captured.

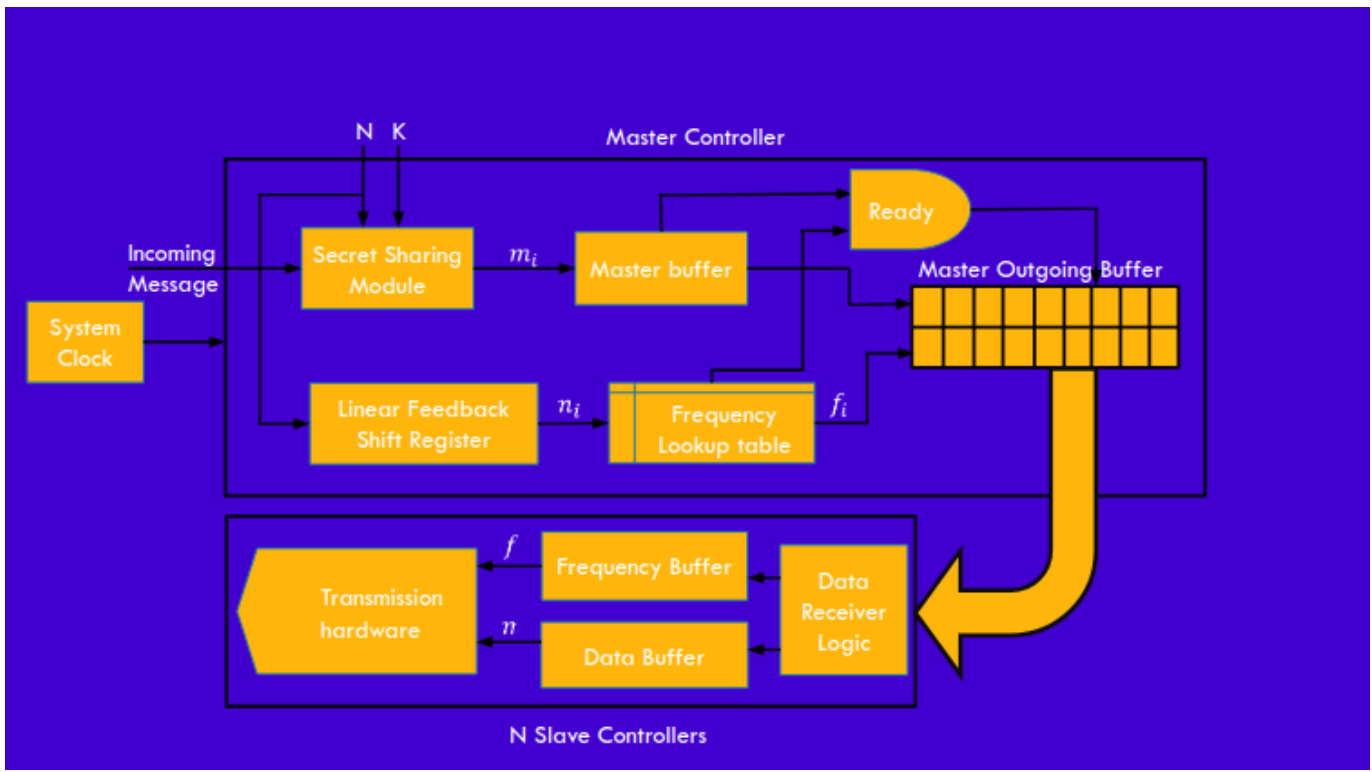

Fig. 7. Prototype Design of Proposed Method

$$P_1(e) = \sum_{i=1}^{\infty} \frac{1}{100000_i} \quad (1)$$

where i is the number of packets being captured.

With secret sharing, let the probability of a packet being captured be $P_2$. Entire information is revealed if and only if all k packets are captured, where k is the number of parts data is divided into. So the only case when information capture occurs is if all k partitions of the data are captured, which is accounted for in $P_2$

$$P_2(e) = \frac{1}{\binom{100000}{n}}$$

Clearly we can see that $P_1 > P_2$ using Theorem 1. In terms of security it can be calculated for k = 10, $P_1(e) = 2^{-18}$ whereas $P_2(e) = 2^{-160}$. This implies that with secret sharing the system exhibits a very low probability of data capture as compared to the system that does not use secret sharing. This means the system becomes more secure. **Theorem 1**: Let **n** be the size of a set, say $F$ for a given event and **k** be the maximum number of elements of $F$ allowed to be chosen. Then consider two cases:

**Case 1**: You are allowed to pick exactly k objects out of n objects. The number of possible ways are $W_1 = \binom{n}{k}$ (i)

**Case 2**: You are allowed to pick at most k objects and at least 1 object out of n objects. The number of possible ways are $W_2 = \sum_{i=1}^{k} \binom{n}{i}$ ......(ii)

Subtracting equation(ii) from equation(i) we get $W_2 - W_1 = \sum_{i=1}^{k-1} \binom{n}{i}$ which is clearly greater than zero. This means $W_2 \geq W_1$, which in turn states $\frac{1}{W_2} \leq \frac{1}{W_1}$

Hence it has been proved that secret sharing increases the security of the system by a huge exponent. Due to multi channel communication and secret sharing, this system is highly resilient and less prone to security breaches.

## VI. COMPARISON BETWEEN EXITING AND PROPOSED SYSTEM

In choosing a wireless technology the following constraints are to be considered. [38] [39]
- Bandwidth
- Security
- Cost
- Energy Consumption

To compare with the existing secure communication mechanism with our proposed system we have chosen Bandwidth, Security, Cost, Energy Consumption [38] [39] and Resource Optimisation in the comparison matrix. [ Table-III].

In the comparison matrix [Table-III] it has been depicted that the bandwidth, cost and energy consumption are low whereas the security and resource optimisation are high in case of existing crytographic secure communication mechanism. On the other hand bandwidth, cost and energy consumption are high whereas the security and resource optimisation are low in our proposed system. From the above matrix analysis it is seen that the acceptability of our proposed system is high with respect to all matrix parameters.

## VII. CONCLUSION

The proposed prototype has been successfully implemented with proposed mechanism. Based on different wireless com-munication attributes it has been observed that the proposed multi channel secure transmission mechanism is robust and also resilient compared to the existing secure communication mechanism. The communication channel overhead and the device level memory utilisation have been optimised in the system. The implemented prototype can be operated with any other conventional secure communication method/s to obtain multi layer security. For this heterogeneous nature the prototype shall be widely accepted and used as a novel secure communication system for radio link transmission.

TABLE III
COMPARISON MATRIX FOR EXISTING AND PROPOSED SYSTEM

| Matrix Parameter | Cryptography System | Proposed System | Acceptability of Proposed System |
|---|---|---|---|
| Bandwidth | High | Low | High |
| Security | Low | High | High |
| Cost | High | Low | High |
| Energy Consumption | High | Low | High |
| Resource Optimisation | Low | High | High |